\documentclass [aps,pra,twocolumn,showpacs]{revtex4}
\newcommand{\ket}[1]{|#1\rangle} 
\newcommand{\bra}[1]{\langle #1|} 
\newcommand{\Tr}{\text{Tr}} 
\usepackage{graphicx}
\begin{document}
\title{Adiabatic geometric phases in hydrogenlike atoms}
\author{Erik Sj\"oqvist$^{1}$\footnote{Electronic address: 
eriks@kvac.uu.se}, X. X. Yi$^{2}$\footnote{Electronic address: 
yixx@dlut.edu.cn}, and Johan {\AA}berg$^{1}$\footnote{Electronic address: 
johan.aberg@kvac.uu.se}}
\affiliation{$^{1}$Department of Quantum Chemistry, Uppsala University, 
Box 518, Se-751 20 Uppsala, Sweden \\ 
$^{2}$Department of Physics, Dalian University of Technology, 
Dalian 116024, China}
\date{\today}
\begin{abstract} 
We examine the effect of spin-orbit coupling on geometric phases in 
hydrogenlike atoms exposed to a slowly varying magnetic field. The 
marginal geometric phases associated with the orbital angular momentum 
and the intrinsic spin fulfill a sum rule that explicitly relates them 
to the corresponding geometric phase of the whole system. The marginal 
geometric phases in the Zeeman and Paschen-Back limit are analyzed. 
We point out the existence of nodal points in the marginal phases 
that may be detected by topological means.  
\end{abstract}
\pacs{03.65.Vf} 
\maketitle
Imagine a quantal spin evolving under influence of a magnetic field so
that the initial and final states of the spin coincide. Cyclic
evolution of this kind results in a phase factor divisible into a
dynamic part and a part that only depends upon the global geometry
associated with the evolution of the spin. The latter is the geometric
phase, first delineated by Berry \cite{berry84} in the adiabatic
case. This adiabatic geometric phase is proportional to the solid
angle enclosed by the direction of a slowly changing magnetic field
and where the proportionality factor is given by the spin projection
quantum number. The geometric phase structure for this system
resembles exactly that of a charged particle in a magnetic monopole
field. A similar result in the special case of spin$-\frac{1}{2}$ was
subsequently found for nonadiabatic evolution
\cite{aharonov87}, in case of which the solid angle is the area
enclosed on the Bloch sphere.

These results opened up the possibility to study magnetic monopole
structures in the laboratory; a fact that has triggered considerable
interest in the geometric phase for quantal systems carrying angular 
momentum. Extensions of the spin-monopole problem to systems 
consisting of several coupled angular momenta have been theoretically 
put forward
\cite{tang95,sjoqvist00a,ekert00,tong03a,bertlmann04,carollo05,kay05} 
and experimentally implemented \cite{jones00,du03}. In particular, the 
issue concerning the relation between the overall geometric phase and 
the geometric phases of the subsystems has been addressed for 
adiabatically evolving pairs of uniaxially coupled spin$-\frac{1}{2}$ 
\cite{yi04a,yi04b}.

Coupling generally leads to entangled multiparticle systems, which
implies that the marginal states of the concomitant subsystems are
mixed. Geometric phases for mixed states take the form of weighted
averages of geometric phase factors, with weight factors given by the
time-independent \cite{sjoqvist00b} or time-dependent \cite{tong04}
eigenvalues of the corresponding marginal states.

In this paper, we address the issue of coupled angular momenta in
terms of hydrogenlike atoms coupled to a slowly varying
magnetic field. For such systems, there is a natural bipartite
decomposition of the total angular momentum into two subsystems
consisting of the orbital part (L) and the intrinsic spin part (S),
exposed to spin-orbit (LS) coupling. We wish to examine the effect of
the LS coupling on the overall geometric phase as well as on those
pertaining to the two subsystems.

Consider a hydrogenlike atom driven by a uniform magnetic field ${\bf
B} = B_0 {\bf n}$ with $B_0$ the nonzero magnetic field strength and
${\bf n} = (\sin \theta \cos \phi , \sin \theta \sin \phi,\cos\theta)$, 
$\theta$ and $\phi$ being slowly varying parameters. Let ${\bf L}$ and 
${\bf S}$ be the orbital angular momentum and intrinsic spin, respectively. 
The spin-orbit Hamiltonian reads \cite{sakurai94}
\begin{eqnarray} 
H_{\bf n} & = & g {\bf n} \cdot ({\bf L} + 2{\bf S}) + 
2{\bf L} \cdot {\bf S} 
\nonumber \\ 
 & = & U_L (\theta,\phi) U_S (\theta,\phi) H_z 
U_L^{\dagger} (\theta,\phi) U_S^{\dagger} (\theta,\phi) 
\nonumber \\ 
 & = & U_J (\theta,\phi) H_z 
U_J^{\dagger} (\theta,\phi) , 
\label{eq:soham}
\end{eqnarray} 
where we may choose ($\hbar = 1$ from now on) 
\begin{eqnarray} 
U_X (\theta,\phi) & = & 
e^{-i\phi X_z} e^{-i\theta X_y} e^{i\phi X_z} , \ X=L,S,J ,  
\end{eqnarray} 
${\bf J} = {\bf L} + {\bf S}$ being the total angular momentum. Here,
$g$ is the Zeeman-LS strength ratio, $H_z$ is the Hamiltonian at the
north pole ${\bf n}=(0,0,1)$ of the parameter sphere, and
$[H_z,J_z]=0$; the latter implying that the eigenvectors of $H_z$ 
can be labeled by the eigenvalues $\mu$ of $J_z$. 

$H_z$ is block-diagonalizable in one- and two-dimensional blocks with
respect to the product basis with elements $\ket{l,m}\ket{\frac{1}{2}, 
\pm \frac{1}{2}} \equiv \ket{l,m} \ket{\pm}$ being the common 
eigenvectors of ${\bf L}^2,L_z,{\bf S}^2,S_z$. Each block may be 
labeled by the eigenvalue $\mu=-l-\frac{1}{2},-l+\frac{1}{2},
\ldots, l+\frac{1}{2}$ of $J_z$. The two extremal subspaces characterized
by $|\mu|=l+\frac{1}{2} \equiv \mu_{\textrm{\scriptsize{e}}}$ are
one-dimensional corresponding to the two product vectors
\begin{eqnarray} 
\ket{\psi^{(l;\pm \mu_{\textrm{\scriptsize{e}}}})} = 
\ket{l,\pm l} \ket{\pm}. 
\label{eq:extevect} 
\end{eqnarray} 
The remaining blocks are two-dimensional, each of which corresponding 
to the vectors $\ket{l,m=\mu-\frac{1}{2}} \ket{+}, \ket{l,m+1=
\mu+\frac{1}{2}} \ket{-}$, $|\mu| < l+\frac{1}{2}$. For each 
such $\mu$, the corresponding two-dimensional Hamiltonian 
suboperator has the form 
\begin{eqnarray} 
H_z^{(l;\mu)} & = & E^{(l;\mu)} I^{(l;\mu)} + \Delta E^{(l;\mu)} 
\left( \sin \alpha^{(l;\mu)} \ \sigma_x^{(l;\mu)} \right. 
\nonumber \\ 
 & & \left. + \cos \alpha^{(l;\mu)} \ \sigma_z^{(l;\mu)} \right) , 
\end{eqnarray} 
with $I^{(l;\mu)}$, $\sigma_x^{(l;\mu)}$, and $\sigma_z^{(l;\mu)}$ the 
standard unit and Pauli operators acting on the relevant subspace. 
Furthermore 
\begin{eqnarray} 
E^{(l;\mu)} & = & g\mu - \frac{1}{2},   
\nonumber \\ 
\Delta E^{(l;\mu)} & = & \frac{1}{2} 
\sqrt{g^2 + 4g \mu + 
\big( 2l+1 \big)^2} , 
\nonumber \\ 
\cos \alpha^{(l;\mu)} & = & 
\frac{2\mu + g}{\sqrt{g^2 + 4g \mu + \big( 2l+1 \big)^2}} ,  
\end{eqnarray}
in terms of which the energy eigenvalues read $E_{\pm}^{(l;\mu)} = 
E^{(l;\mu)} \pm \Delta E^{(l;\mu)}$ with corresponding entangled 
eigenvectors  
\begin{eqnarray} 
\ket{\psi_+^{(l;\mu)}} & = & 
\cos \big( \frac{1}{2} \alpha^{(l;\mu)} \big) \ 
\ket{l,\mu-\frac{1}{2}} \ket{+} 
\nonumber \\ 
 & & + 
\sin \big( \frac{1}{2} \alpha^{(l;\mu)} \big) \ 
\ket{l,\mu+\frac{1}{2}} \ket{-} , 
\nonumber \\ 
\ket{\psi_-^{(l;\mu)}} & = & 
-\sin \big( \frac{1}{2} \alpha^{(l;\mu)} \big) \ 
\ket{l,\mu-\frac{1}{2}} \ket{+} 
\nonumber \\ 
 & & + 
\cos \big( \frac{1}{2} \alpha^{(l;\mu)} \big) \ 
\ket{l,\mu+\frac{1}{2}} \ket{-} .  
\label{eq:evect}
\end{eqnarray}
The $g$-dependence of the eigenvectors is due to the fact that 
the Zeeman and LS term do not commute. 

With the above choice of rotation operators, we have 
$U_X (0,\phi) = I$ and $U_X (\pi,\phi) = e^{-i\pi X_y} e^{2i\phi
X_z}$. This entails that the corresponding energy eigenvectors cannot
be unique simultaneously at the north and south pole. For example, by
choosing the phase of the eigenvectors $\ket{\psi^{(l;\mu)}}$ of $H_z$
to be independent of $\phi$, as in Eqs. (\ref{eq:extevect}) and
(\ref{eq:evect}), the resulting instantaneous eigenvectors $U_X
(\theta,\phi) \ket{\psi^{(l;\mu)}}$ are unique at the north pole but 
yields a singular gauge potential at the south pole. For the same 
reference eigenvectors, one may move this singularity to the north pole 
by instead choosing the rotation operators $\widetilde{U}_X (\theta,\phi)
= e^{-i\phi X_z} e^{-i\theta X_y} e^{-i\phi X_z}$. On the other 
hand, the section \cite{wu75} $\{ U_X (\theta,\phi),\theta \in [0,\pi);
\widetilde{U}_X (\theta,\phi),\theta \in (0,\pi]\}$ is globally 
well-defined for any choice of eigenvectors of $H_z$. This
single-valued section captures the monopole structure corresponding to
the $2j+1$ fold degeneracy at $g=0$, $j$ being the eigenvalue of ${\bf
J}^2$.

We now compute the adiabatic geometric phases for the atom and its
subsystems L and S under the assumption $g\neq 0$. Let us start with 
the extremal states $\mu = \pm
\mu_{\textrm{\scriptsize{e}}}$. We note that
\begin{eqnarray} 
\ket{\psi^{(l;\pm \mu_{\textrm{\scriptsize{e}}})}; \theta,\phi} & = & 
U_L (\theta,\phi) \ket{l,\pm l} U_S (\theta,\phi) 
\ket{\pm} 
\end{eqnarray}
are product eigenvectors of $H_{\bf n}$. Assume that the external
magnetic field slowly traverses a loop $\mathcal{C}$ such that the 
adiabatic approximation is valid. Then, the adiabatic geometric phase 
becomes 
\begin{eqnarray} 
\Gamma_J^{(l;\pm \mu_{\textrm{\scriptsize{e}}})} [\mathcal{C}] = 
\mp \mu_{\textrm{\scriptsize{e}}} \Omega  
\label{eq:cs}
\end{eqnarray}
with $\Omega$ the solid angle enclosed by the loop. From the product
form of the extremal states, we obtain the corresponding marginal
geometric phases for L and S as $\mp l\Omega$ and $\mp \frac{1}{2}
\Omega$, respectively, which are $g$-independent and sum up to
$\Gamma_J^{(l;\pm \mu_{\textrm{\scriptsize{e}}})} [\mathcal{C}]$ since 
$\mu_{\textrm{\scriptsize{e}}} = l+\frac{1}{2}$.

Next we compute the adiabatic geometric phases for $|\mu| <
l+\frac{1}{2}$. The eigenvectors of the instantaneous Hamiltonian 
$H_{{\bf n}}$ take the form 
\begin{eqnarray} 
\ket{\psi_{\pm}^{(l;\mu)};\theta,\phi} & = & 
U_J(\theta,\phi) \ket{\psi_{\pm}^{(l;\mu)}} 
\nonumber \\ 
 & = & U_L(\theta,\phi) U_S(\theta,\phi) \ket{\psi_{\pm}^{(l;\mu)}} .  
\end{eqnarray} 
We obtain the $g$-independent pure state geometric phase as
\begin{eqnarray} 
\Gamma_{J,\pm}^{(l;\mu)} [\mathcal{C}] = -\mu \Omega , 
\label{eq:totalgp}
\end{eqnarray}
which follows directly from the fact that the energy eigenvectors 
$\ket{\psi_{\pm}^{(l;\mu)}; \theta,\phi}$ are also eigenvectors of 
${\bf n} \cdot {\bf J}$ both with the eigenvalue $\mu$.  The marginal 
states read  
\begin{eqnarray} 
\rho_{L,\pm}^{(l;\mu)} (\theta,\phi) & = & 
\Tr_S \ket{\psi_{\pm}^{(l;\mu)};\theta,\phi} 
\bra{\psi_{\pm}^{(l;\mu)};\theta,\phi} 
\nonumber \\ 
 & = & U_L(\theta,\phi) \rho_{L,\pm}^{(l;\mu)} 
U_L^{\dagger} (\theta,\phi) , 
\nonumber \\ 
\rho_{S,\pm}^{(l;\mu)} (\theta,\phi) & = & 
\Tr_L \ket{\psi_{\pm}^{(l;m)};\theta,\phi} 
\bra{\psi_{\pm}^{(l;\mu)};\theta,\phi} 
\nonumber \\ 
 & = & U_S(\theta,\phi) \rho_{S,\pm}^{(l;\mu)} 
U_S^{\dagger}(\theta,\phi) , 
\end{eqnarray} 
with 
\begin{eqnarray} 
\rho_{L,\pm}^{(l;\mu)} & = & 
\Tr_S \ket{\psi_{\pm}^{(l;\mu)}} \bra{\psi_{\pm}^{(l;\mu)}} 
\nonumber \\ 
 & = & \frac{1}{2} \left( 1 \pm \cos \alpha^{(l;\mu)} \right) 
\ket{l,\mu-\frac{1}{2}} \bra{l,\mu-\frac{1}{2}} 
\nonumber \\ 
 & & + \frac{1}{2} \left( 1 \mp \cos \alpha^{(l;\mu)} \right)
\ket{l,\mu+\frac{1}{2}} \bra{l,\mu+\frac{1}{2}} , 
\nonumber \\ 
\rho_{S,\pm}^{(l;\mu)} & = & 
\Tr_L \ket{\psi_{\pm}^{(l;\mu)}} \bra{\psi_{\pm}^{(l;\mu)}} 
\nonumber \\ 
 & = & 
\frac{1}{2} \left( 1 \pm \cos \alpha^{(l;\mu)} \right) 
\ket{+} \bra{+} 
\nonumber \\ 
 & & + \frac{1}{2} \left( 1 \mp \cos \alpha^{(l;\mu)} \right) 
\ket{-} \bra{-} . 
\end{eqnarray} 
Since the marginal density operators $\rho_{L,\pm}^{(l;\mu)}$ and 
$\rho_{S,\pm}^{(l;\mu)}$ evolve unitarily under $U_L$ and $U_S$, 
respectively, it follows that the marginal geometric phases can 
be computed using the approach in Ref. \cite{sjoqvist00b}. Explicitly, 
for a magnetic field whose direction traces out a loop $\mathcal{C}$, 
this yields 
\begin{widetext}
\begin{eqnarray} 
\exp \left( i\Gamma_{L,\pm}^{(l;\mu)} \big[ \mathcal{C};g \big] \right) & = &  
\Phi \Big[ \left( 1 \pm \cos \alpha^{(l;\mu)} \right) 
e^{-i(\mu-\frac{1}{2})\Omega} + 
\left( 1 \mp \cos \alpha^{(l;\mu)} \right) 
e^{-i(\mu+\frac{1}{2})\Omega} \Big] 
\nonumber \\ 
 & \Rightarrow & \Gamma_{L,\pm}^{(l;\mu)} \big[ \mathcal{C};g \big] =  
-\mu \Omega \pm \arctan \left( \cos \alpha^{(l;\mu)} 
\tan \frac{\Omega}{2} \right) ,  
\nonumber \\ 
\exp \left( i\Gamma_{S,\pm}^{(l;\mu)} 
\big[ \mathcal{C};g \big] \right) & = &   
\Phi \Big[ \left( 1 \pm \cos \alpha^{(l;\mu)} \right) e^{-i\Omega/2} + 
\left( 1 \mp \cos \alpha^{(l;\mu)} \right) e^{i\Omega/2} \Big] 
\nonumber \\ 
 & \Rightarrow & \Gamma_{S,\pm}^{(l;\mu)} \big[ \mathcal{C};g \big] =  
\mp \arctan \left( \cos \alpha^{(l;\mu)} 
\tan \frac{\Omega}{2} \right) . 
\label{eq:marginalgp}
\end{eqnarray}
\end{widetext}
Here, $\Phi [z]=z/|z|$ for any nonzero complex number $z$. Notice 
that the above marginal geometric phases are $g$-dependent through $\cos
\alpha^{(l;\mu)}$. They obey the symmetry $\Gamma_{X,\pm}^{(l;-\mu)} 
\big[ \mathcal{C};-g \big] = -\Gamma_{X,\pm}^{(l;\mu)} 
\big[ \mathcal{C};g \big]$, $X=L,S$, which is expected since the change 
$(\mu,g,\Omega)\rightarrow (-\mu,-g,\Omega)$ is physically equivalent
to reversing the orientation of the loop $\mathcal{C}$. Furthermore, by
comparing Eqs. (\ref{eq:totalgp}) and (\ref{eq:marginalgp}), it
follows that the marginal geometric phases of the L and S subsystems
fulfill the sum rule 
\begin{eqnarray} 
\Gamma_{L,\pm}^{(l;\mu)} \big[ \mathcal{C};g \big] + 
\Gamma_{S,\pm}^{(l;\mu)} \big[ \mathcal{C};g \big] = 
\Gamma_{J,\pm}^{(l;\mu)} [\mathcal{C}] 
\end{eqnarray}
that explicitly relates them to the corresponding geometric phases for
the pure entangled states. 

Let us now consider the extreme cases $|g| \gg 1$ (Paschen-Back regime 
\cite{remark}) and $0<|g| \ll 1$ (Zeeman regime). In the Paschen-Back 
limit, we have $\cos \alpha^{(l;\mu)} \approx 1$, which implies
\begin{eqnarray}
\Gamma_{S,\pm}^{(l;\mu)} \big[ \mathcal{C};g \big] & \approx & 
\mp \frac{1}{2} \Omega , 
\nonumber \\ 
\Gamma_{L,\pm}^{(l;\mu)} \big[ \mathcal{C};g \big] & \approx & 
-\left( \mu \mp \frac{1}{2} \right) \Omega . 
\end{eqnarray} 
These phases are those of the pure vectors $U_S (\theta,\phi)\ket{\pm}$ 
and $U_L (\theta,\phi)\ket{l,\mu \pm \frac{1}{2}}$, respectively, as 
expected as the LS term in $H_{{\bf n}}$ is negligible in this limit.  
The Zeeman condition $0<|g|\ll 1$ yields $\cos \alpha^{(l;\mu)} \approx 
\mu /(l+\frac{1}{2})$, leading to    
\begin{eqnarray}
\Gamma_{S,\pm}^{(l;\mu)} \big[ \mathcal{C};g \big] & \approx & \mp 
\arctan \left( \frac{\mu}{l+\frac{1}{2}} \tan \frac{\Omega}{2} \right) , 
\nonumber \\  
\Gamma_{L,\pm}^{(l;\mu)} \big[ \mathcal{C};g \big] & \approx & -\mu \Omega 
\pm \arctan \left( \frac{\mu}{l+\frac{1}{2}} \tan \frac{\Omega}{2} \right) .  
\end{eqnarray}
From this we can conclude that the marginal geometric phases may in
general not be small for $0<g \ll 1$, contrary to the cases discussed
in Refs. \cite{yi04a,yi04b}, where all geometric phases were found to be
quenched in this regime due to the uniaxial coupling term. The reason
for this quenching effect in the uniaxial case is that the coupling
term defines a fixed preferred quantization axis that makes the
eigenstates essentially unaffected by a weak magnetic field. In the LS
case, though, no particular direction in space is singled out by the
spherically symmetric coupling term and the quantization axis of the
instantaneous eigenstates still coincides with the direction of the
applied magnetic field. This feature is true no matter how small $g$
is as long as it is nonzero. On the other hand, if $g=0$, then the
magnetic field decouples from the atom and no change in the atomic
eigenstates can take place when the direction of the magnetic field
varies. Thus, $g=0$ is a singular point in the sense that the
geometric phases become independent of the enclosed solid angle
$\Omega$ of the magnetic field. It should be noted that the same
singular behavior is present for the standard case \cite{berry84} of a
single spin in a slowly rotating magnetic field. In physically
realistic scenarios, though, it is reasonable to expect that the
singularity at $g=0$ becomes invisible as the atom is increasingly
exposed to noise and decoherence effects.

\begin{figure}[htb]
\includegraphics[width = 8 cm]{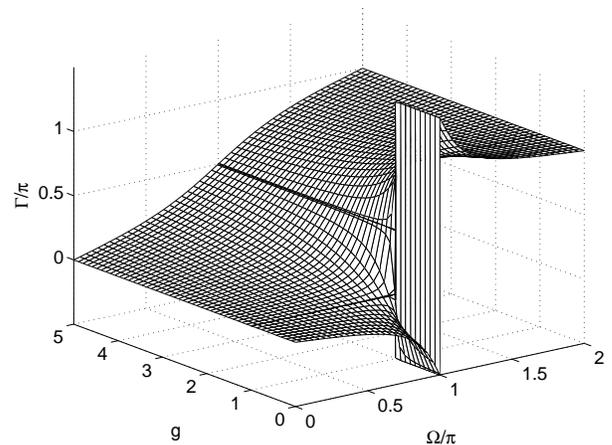}
\caption{\label{fig:1} Adiabatic geometric phase 
$\Gamma_{S,-}^{(2;-\frac{1}{2})} \big[ \mathcal{C};g \big]$ of the
intrinsic spin as a function of the dimensionless Zeeman-LS coupling
strength ratio $g$ and solid angle $\Omega$ enclosed by the loop
$\mathcal{C}$ of the magnetic field.}
\end{figure}

The geometric phase $\Gamma_{S,-}^{(2;-\frac{1}{2})} \big[ 
\mathcal{C};g \big]$ of the intrinsic spin as a function of 
coupling strength $g$ and solid angle $\Omega$ is shown in 
Fig. \ref{fig:1}. Notice in particular that this graph confirms the 
expected asymptotic behavior in the Paschen-Back limit $|g| \gg 1$. 

The marginal density operators are degenerate when $\cos
\alpha^{(l;\mu)}=0$, which happens along the line $g=-2\mu$ in the
space spanned by $(\Omega,g)$. For $\mu = - \frac{1}{2}$ this line 
occurs at $g=1$, as indicated in Fig. \ref{fig:1}. While the cyclic 
geometric phase of the whole system is well-defined at these lines, 
the corresponding marginal geometric phases are undefined there 
\cite{sjoqvist00b}. Furthermore, the visibilities of the subsystems, 
defined as \cite{sjoqvist00b} 
\begin{eqnarray} 
\mathcal{V}_{\pm,S}^{(l;\mu)} \big[ \mathcal{C};g \big] & = & 
\mathcal{V}_{\pm,L}^{(l;\mu)} \big[ \mathcal{C};g \big] 
\nonumber \\ 
 & \equiv & \frac{1}{2} \left| \left( 1 \pm \cos \alpha^{(l;\mu)} \right) 
e^{-i\Omega/2} \right. 
\nonumber \\ 
 & & + \left. \left( 1 \mp \cos \alpha^{(l;\mu)} \right) 
e^{i\Omega/2} \right| , 
\end{eqnarray}
reduces to 
\begin{eqnarray} 
\mathcal{V}_{\pm,S}^{(l;\mu)} \big[ \mathcal{C};g \big] =  
\mathcal{V}_{\pm,L}^{(l;\mu)} \big[ \mathcal{C};g \big] = 
\left| \cos \frac{\Omega}{2} \right| 
\end{eqnarray} 
at the points where the marginal geometric phases are undefined. The 
marginal visibilities vanish at their common nodal points $\Omega = 
(2p+1) \pi$, $p$ integer, which is manifested as a jump at 
$(\Omega,g)=(\pi,1)$ in Fig. \ref{fig:1}. These nodal points 
can be detected topologically by considering loops in the space 
spanned by $(\Omega,g)$. By continuously monitoring the marginal 
phases along a loop, a resulting $2\pi$ phase shift signals the 
existence of such a nodal point \cite{bhandari02}. For example, 
as indicated in Fig. \ref{fig:1}, by traversing a loop in the 
counterclockwise (clockwise) direction such that it encloses 
the singular point at $(\Omega,g)=(\pi,1)$ once, we end up at a 
phase shift $2\pi$ ($-2\pi$).  

In conclusion, we have computed adiabatic geometric phases in
hydrogenlike atoms coupled to a slowly varying magnetic field. We have
shown that while the total geometric phase is independent of the
strength of the spin-orbit coupling, this is not the case for the
corresponding marginal phases. It turns out, though, that the latter
phases sum up to the pure state geometric phase of the whole
system. Further consideration as to the generality of this sum rule
for other systems of coupled angular momenta seems pertinent. We have
examined the marginal geometric phases in the Zeeman and Paschen-Back
limit. Finally, we have pointed out the
existence of nodal points where the marginal geometric phases become
undefined and we have argued that these points may be detected by
topological means.
\vskip 0.3 cm 
X.X.Y. acknowledges financial support by NSF of China (Project 
No. 10305002).
 

\begin{thebibliography}{99}
\bibitem{remark} In this context, it is important to notice that the 
diamagnetic term, typically of the order $g^2$, in the exact
Hamiltonian has been neglected in Eq. (\ref{eq:soham}). To
justify this approximation in the Paschen-Back $|g|\gg 1$ regime, it
therefore becomes important that $g$ is not too large (see, e.g.,
Ref. \cite{tang95} for some numerical estimates in the hydrogen atom
case).
\end{thebibliography}
\end{document}